# Using the Value of Information (VoI) Metric to Improve Sensemaking

Mark Mittrick[1], John Richardson[1], Derrik E. Asher[1], Alex James[2], Timothy Hanratty[1]

**1 U.S. Army Research Laboratory**
Computational Information Sciences Directorate
Multilingual Computing Branch
Aberdeen Proving Ground, Maryland 21005

**2 CUBRC**
4455 Genesee Street, Suite 106
Buffalo, NY  14225

# Abstract

Sensemaking is the cognitive process of extracting information, creating schemata from knowledge, making decisions from those schemata, and inferring conclusions. Human analysts are essential to exploring and quantifying this process, but they are limited by their inability to process the volume, variety, velocity, and veracity of data. Visualization tools are essential for helping this human-computer interaction. For example, analytical tools that use graphical link-node visualization can help sift through vast amounts of information. However, assisting the analyst in making connections with visual tools can be challenging if the information is not presented in an intuitive manner.

Experimentally, it has been shown that analysts increase the number of hypotheses formed if they use visual analytic capabilities. Exploring multiple perspectives could increase the diversity of those hypotheses, potentially minimizing cognitive biases. In this paper, we discuss preliminary research results that indicate an improvement in sensemaking over the traditional link-node visualization tools by incorporating an annotation enhancement that differentiates links connecting nodes. This enhancement assists by providing a visual cue, which represents the perceived value of reported information. We conclude that this improved sensemaking occurs because of the removal of the limitations of mentally consolidating, weighing, and highlighting data. This study aims to investigate whether line thickness can be used as a valid representation of VoI.

# 1. Introduction

In 1993, Russell et al. introduced the concept of sensemaking to the human-computer interaction community [1]. They identified it as a common activity in analysis, involving the process of searching for a representation and encoding data in that representation to answer task questions. Following that, researchers sought to begin applying sensemaking to visualization capabilities.

Around the same time sensemaking was introduced, Larkin and Simon (1987) suggested analysts' could spot anomalies and other patterns, if the burden of mentally consolidating information is minimized [2]. This statement identifies the underlying assumption or premise of why many analytical and visualization tools are useful – they enable users to gain insights that are otherwise obscured. Visualization tools affect the amount of cognition needed to solve problems by reducing the difficulty level of finding and comparing data.

A prominent visualization tool is the link-node diagram. This visualization technique combines nodes with connecting links to create a network of associated nodes [3]. In testing whether the human-interactive aspect of link-node analysis could be improved, Ware and Bobrow (2005) researched techniques highlighting a small number of nodes to determine whether a large network could be displayed, while maintaining the effective visualization power of a small link-node diagram [4]. Their research combined visual highlighting and motion-cues to emphasize a small numbers of nodes, and compared the effectiveness of the visual cues to baseline results. Their results showed that analysts could answer questions with undirected graphs of less than a hundred nodes, compared to a baseline level of error. However, as the undirected graph grew larger, performance approached chance levels. When highlighting was introduced, error levels dropped substantially. In essence, it was demonstrated that pre-attentive cues are effective within the context of large and complex link-node visualizations. In another study, it was found that using a weighting scheme that displayed a link's length in proportion to its weight improved comprehensibility of link-node graphs representing webpage similarity data [5]. Together, these studies provide sufficient evidence for investigating the efficacy for line thickness to improve sensemaking in link-node diagrams.

## Value of Information

Value of Information (VoI) is a metric that computes a likelihood of applicability based on metadata of recorded information. Specifically, the Value of Information (VoI) metric combines source reliability, likelihood of data being true, and timeliness with respect to mission [6], [7]. The current pilot study tests the perceived value of line thickness within a link-node visualization framework, to quantitatively assess the performance gains when utilizing link-node density as an independent variable. If it is determined, that line thickness helps reduce the mental burden on analysts, then it is suitable to be used in VoI paradigms.

## Crowdsourcing

R. Cialdini and M. Trost (1998) described crowdsourcing as a process of outsourcing difficult to answer questions to a crowd of individuals [8]. The power of crowdsourcing comes from the concept "wisdom of the crowd," which indicates that a large number of individuals estimating some phenomena will produce an averaged estimate that is as good as or often better than that of an expert [9], [10]. An explanation for this phenomenon is that noise inherently exists in estimates and an average over a large amount of these noisy estimates results in a reduction in the overall noise, abiding by the law of large numbers in probability theory [11], [12]. Given the power of the crowdsourcing method, it is an ideal choice for examining whether or not VoI helps improve sensemaking.

The work presented in this paper shows how crowdsourcing informs the VoI paradigm with a non-analyst population, utilizing response time and performance accuracy determined from degree of centrality.

## 2. Methods

In this experiment, Amazon Mechanical Turk (MTurk) was the crowdsourcing platform used to collect the data from 303 participants. A simple computerized task required subjects to review a link-node diagram presented to them, and then to select the node they "know most about". Each subject was randomly assigned to one of six conditions, which consisted of three levels (Easy, Medium, Hard) and two groups (VoI and Control).

## Human Subjects

This study falls under the U.S. Army Research Laboratory's (ARL) internal review board (IRB) Exempt Research Determination for Protocol (ARL 17-093). This indicates that the research is exempt from regulations 32 CFR 219. The research is exempt because it falls into the exemption criteria defined by the Common Rule, which states that human subjects cannot be identified by the collected data, and the responses provided by the subjects place them at no risk of criminal or civil liability or be damaging to their financial standing, employment, or reputation.

Upon volunteering to participate in the study, subjects were notified that needed to be familiar with link-node diagrams, and no personally identifiable information would be collected. Successful completion of the experiment would net the subject a quarter (25¢) with a possible additional bonus quarter (25¢) if the subject was able to correctly identify the node they know the most about.

For subjects who qualified, three demographic questions were asked of each. They focused on the subjects Occupation, Age, and Education Level. Exclusionary criteria consisted of the subjects' knowledge of link-node diagrams. If a participant indicated that they were not familiar with link-node diagrams, they program exited (without collecting any data), and the subjects were thanked for their interest in the study.

User bias was minimized by only allowing each subject to participate in the study once. This was accomplished by setting the eligibility criteria to only allow subjects to play one time. If a subject attempted to participate a second time mTurk informed them, that they were no longer eligible.

## Experimental Setup

Once a subject was deemed qualified, they were presented with instructions and an example graph to practice on. The instructions stated that the subject was to imagine that they were an analyst studying a link-node diagram. Each link incident upon a node represents the metric to be maximized. The thickness of the link indicates greater value. The thicker the link the more relevant the node becomes (Figure 1). The subject was required to:

1) Assess the diagram to discern the node with the greatest degree of centrality, which was modulated by line thickness in the VoI cases.
2) Select the node via mouse click thus highlighting the node and indicating that a selection has been made. The selected node as also recorded in a list next to the diagram. If the subject was unhappy with their selection, they were able to press the Reset button in order to begin the selection process over.
3) When the subject was happy with their choice they pressed the Submit button to record their answer.
4) The subject was able to drag the nodes and manipulate the graph in order to optimally assess the degree centrality.

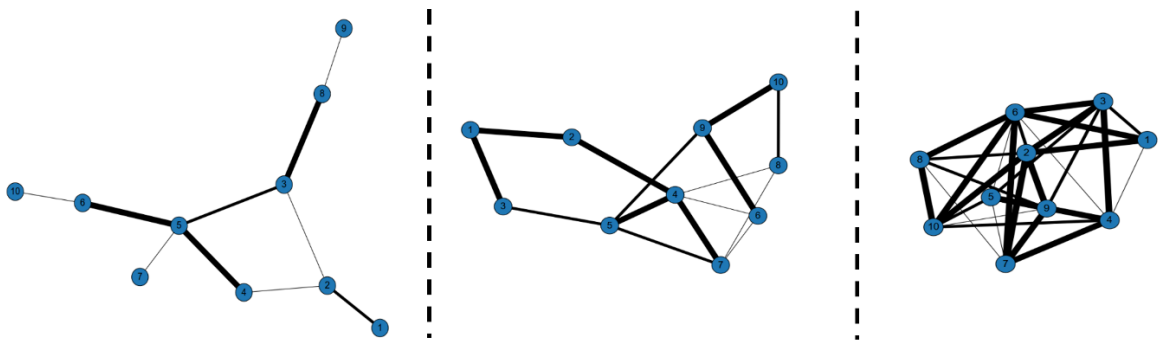

*Figure 1: Link-Node diagrams from experiment showing Easy-VoI (left), Medium-VoI (center), and Hard-VoI (right)*

## Data Analysis Methods

The Jarque-Bera test indicates whether data comes from a Normal distribution with an unspecified mean and standard deviation. The Jarque-Bera test was utilized to confirm that the data was not normally distributed, and therefore, the appropriate analyses required non-parametric tests.

Wilcoxon test is a non-parametric test. The null hypothesis we used for the Wilcoxon test indicates that the distributions of the compared samples are equal. Small p values reject the null and imply that the distributions and medians are not equal.

The Kruskal-Wallis test is equivalent to the Wilcoxon test and was used to confirm the statistical results.

The p-values generated from the Wilcoxon and Kruskal-Wallis tests represent the statistical significance associated with all compared data samples from the respective conditions. P-values were considered significant if below 0.05 and corrections for multiple comparisons were not necessary.

## Conditions

In order to keep the experiment consistent across all conditions, the number of nodes was kept constant with the number of links being varied. In this study, there are six conditions: Easy-VoI, Easy-Control, Medium-VoI, Medium-Control, Hard-VoI, and Hard-Control. The graph density formula calculates the density of the graph given a set of nodes and edges. The graph density formula is:

*Equation 1*

$$D = \frac{2 * E}{N(N-1)}$$

where D is the density of the graph, E is the edges in the graph, and N is the number of nodes in the graph (N = 10 for all graphs). In this pilot study, the density was 22% for Easy (10 Edges), 33% for Medium (15 Edges), and 66% for Hard (30 Edges). The initial threshold for each density level was determined by the perceived level of difficulty and confirmed through preliminary data collected. The easy level threshold (Easy-VoI and Easy-Control) was selected in order to allow the subjects the ability to count the number of edges instead of estimating in order to maximize their potential to select the correct node. Based on the preliminary data collected, the medium and hard levels showed that the subjects were not counting but estimating as their performance decreased with increasing level of difficulty.

## 3. Results

Results from the study show that there are significant statistical differences when comparing certain VoI and Control (non-VoI) graphs using Situational Awareness (SA) and Response Time (RT). SA is the sum of the node's links (i.e., degree centrality). RT is the duration from when a subject first saw the graph to when they submitted their answer. These metrics were used to evaluate whether line thickness can be considered a valid visual representation of VoI.

Each graph has a deterministic value (SA) that provides a quantifiable metric to evaluate the subjects' performance. In the Control condition, the links all have a value of one, whereas in the VoI conditions, each link is weighted (one to three) depending on the thickness of the line. One for the least thick line, two for a medium thick line, and three for the thickest line. For example, a node with three links in the Control conditions has an SA = 3 (one for each link), whereas, that same node in the VoI conditions with medium thickness links has an SA = 6 (3 links x 2 medium thickness).

### Data Analysis

Since the data didn't follow a Normal distribution (confirmed with Jarque-Bera tests), non-parametric tests were utilized. Wilcoxon and Kruskal-Wallis tests were performed pairwise to determine statistical differences between conditions. The study was balanced with approximately equal number of subjects per condition (Table 1).

|         | Easy | Medium | Hard |
|---------|------|--------|------|
| **VoI** | 49   | 44     | 55   |
| **Control** | 53 | 50   | 52   |
| **Total** | 102 | 94    | 107  |

*Table 1: Distribution of subjects across conditions*

The Wilcoxon Rank Sum Test Results for SA (Table 2) show that *Easy-VoI vs. Easy-Control* as well as *Hard-VoI vs. Hard-Control* were found to be statistically significant at the alpha = 0.05 level. This suggests that the compared samples come from different underlying distributions. Thus the Easy and Hard levels show significant improvement in performance of choosing the node with the highest degree centrality. In contrast, *Medium-VoI vs. Medium-Control* was not found to be statistically significant. This evidence suggests that line thickness for the Easy and Hard levels of difficulty are a valid representation of VoI (Figure 2).

| Situational Awareness (SA) | |
|---|---|
| **Wilcoxon** | **p Value** |
| Easy-VoI vs. Easy-Control | 0.0404 * |

| Medium-VoI vs. Medium-Control | 0.6361 |
|---|---|
| Hard-VoI vs. Hard-Control | 0.0056 * |

Table 2: Wilcoxon Rank Sum Test Results for SA. * Statistically significant at the alpha = 0.05 level.

The distribution of data was compared pairwise across the levels of difficulty. In the Easy level, the Easy-VoI condition's median, upper and lower quartiles, and whiskers are all the same, as the majority of the subjects performed perfectly. There were a total of 49 subjects in the Easy-VoI condition (Table 1) of which 40 scored perfect (82%). The nine subjects who chose incorrectly are represented by the three visibly distinct outliers (Figure 2: red plus signs). The nine outlier data points overlap due to Normalized Rank performance collapsing into three distinct values. In the Medium level, the results suggest that there is no statistical significance which can be seen in the middle pairwise comparison in Figure 2. Additional data would be needed in order to fully examine this phenomena. Finally, at the Hard level, we have observed statistical significance with the VoI condition (VoI-Hard) subjects outperforming the control condition subjects (Hard-Control).

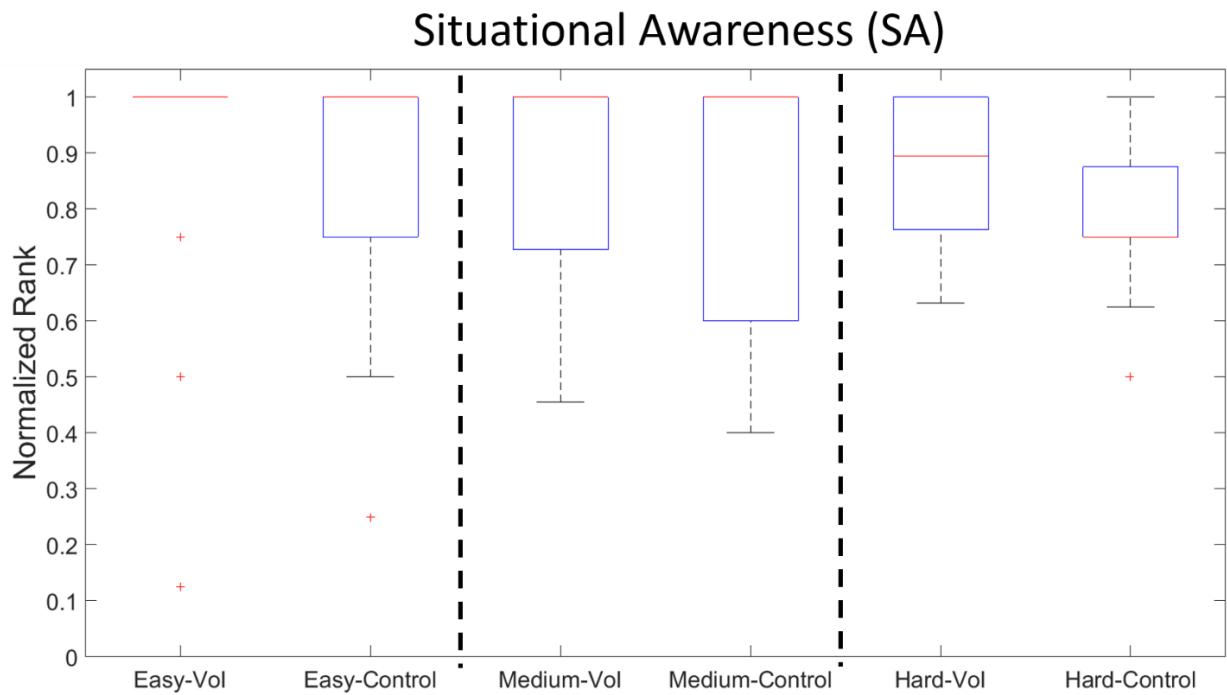

Figure 2: Situational Awareness across conditions: The boxplot shows the distribution of data, with Normalized Rank (y-axis) ranging from 0 - 1 and the 6 conditions (x-axis). The red bars show the medians, the blue bounding boxes show the upper and lower quartiles, the whiskers extend 2.7 standard deviations from the median, and the red plus signs ('+') represent outliers. The vertical dashed lines represent conditions separated by levels of difficulty. Left segment - Easy, middle segment - Medium, and right segment - Hard.

The Wilcoxon Rank Sum Test results for RT (Table 3) show that *Easy-VoI vs. Easy-Control* was found to be statistically significant at the alpha = 0.05 level. This suggests that the compared samples 1) come from different underlying distributions, and 2) that subjects took significantly less time in the VoI case. (Figure 3). In contrast, *Medium-VoI vs. Medium-Control* and *Hard-VoI vs. Hard-Control* were not found to be statistically significant. Additional data may result in significant differences at the Medium and Hard levels.

| Response Time (RT) | |
|---|---|
| Wilcoxon | p Value |
| Easy-VoI vs. Easy-Control | 0.0432 * |
| Medium-VoI vs. Medium-Control | 0.0873 |
| Hard-VoI vs. Hard-Control | 0.7317 |

*Table 3: Wilcoxon Rank Sum Test Results for RT. * Statistically significant at the alpha = 0.05 level.*

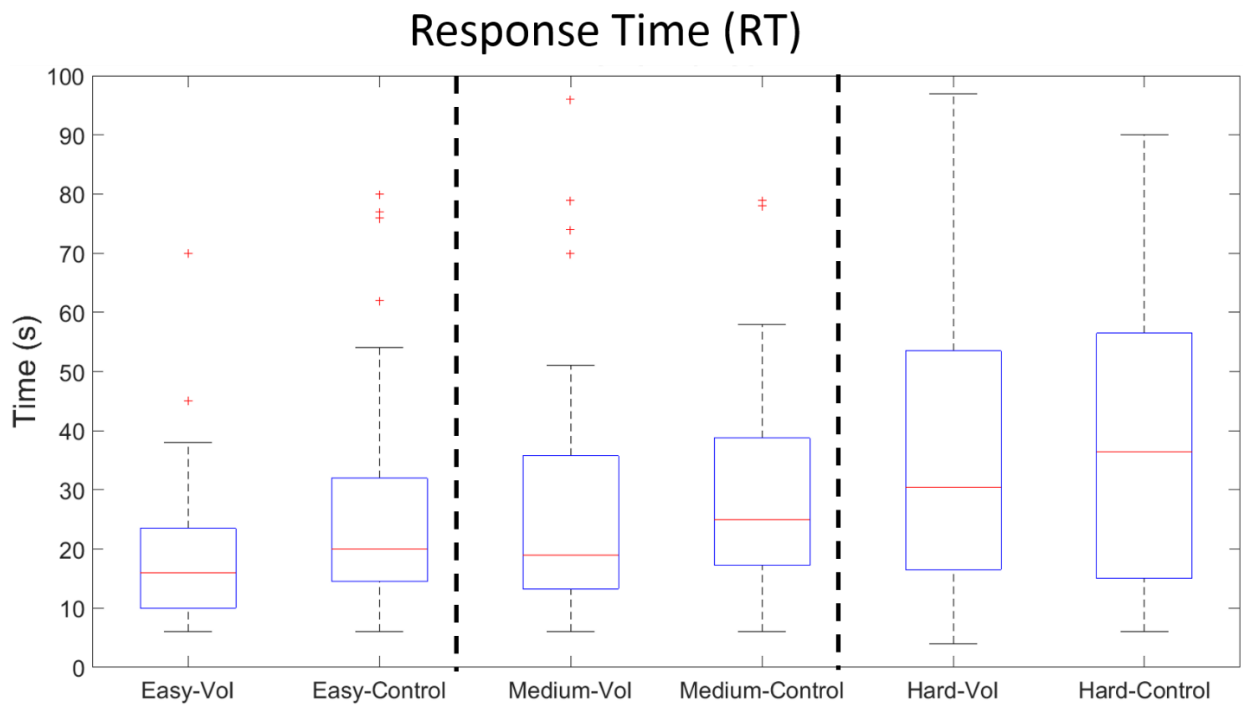

*Figure 3 Response Time across conditions: The boxplot shows the distribution of data, with Time (y-axis) ranging from 0 – 100 (seconds) across the y-axis and the 6 conditions on the x-axis. The red bars show the medians, the blue bounding boxes show the upper and lower quartiles, the whiskers extend 2.7 standard deviations from the median, and the red plus signs ('+') represent outliers. The vertical dashed lines represent conditions separated by levels of difficulty. Left segment - Easy, middle segment - Medium, and right segment - Hard.*

Again, the distribution of data was compared pairwise across the levels of difficulty. In the Easy level, the results suggest that there is statistical significance with the Easy-VoI subjects outperforming the Easy-Control condition subjects. Although no statistical significance was found in the Medium and Hard levels, the medians are trending in a direction consistent with the analysis of SA (Figures 2, 3).

## 4. Discussion

The aim of this study is to determine whether line thickness can be used as a valid representation of VoI. The analysis of Situational Awareness (SA) showed that line thickness for Easy and Hard levels of difficulty, determined through degree centrality, provided a significant improvement in subject performance over the Control conditions (Figure 2). In addition, the analysis of Response Time (RT) showed that subjects performed significantly faster in the Easy level of difficulty due to line thickness (Figure 3: Easy-VoI vs. Easy-Control). Furthermore, the median RT was found to be greater in Control conditions per difficulty level, although not significant for Medium and Hard levels. Together, these results suggest that line thickness might be a viable option to represent VoI.

The results from this experiment are a first step towards improving sensemaking by lessening the mental burden. The ability of VoI to provide a visual cue is paramount to quickly understanding the information presented as well as making important decisions in a quick and timely matter.  Our work shows that using line thickness as a visual cue to the value of a node significantly improves selection performance in the experiment.  Following previous work by Ware and Bobrow (2005) this result shows how cues generated from the perceived value of the data used to create the link node diagram can aid the analyst under certain conditions [4].

Additional data and follow-up experiments are needed to explore how further manipulations to graph density might influence the perceived value of line thickness and help clarify the non-significant findings (Tables 2, 3). We propose a set of follow-up experiments with more participants, different amount of nodes, and additional levels of graph density to determine how best to utilize line thickness and its perceived value. We currently use random number generation to determine the line thickness and placements of the graphs. Currently, we are unable to assess how this impacts our study at this point. In the future, we hope to find a way to normalize this assignment to further control our observations.

ICCRTS – 2017
Category: Human Information Interaction
Title: **Using the Value of Information (VoI) Metric to Improve Sensemaking**